\documentclass[aps,prb,twocolumn,superscriptaddress,floatfix]{revtex4-2}
\pdfoutput=1 
\usepackage{longtable,xfrac}
\usepackage[colorlinks=true,citecolor=blue,linkcolor=blue,breaklinks=true]{hyperref}
\usepackage{amssymb}
\usepackage{graphicx}
\usepackage{amsmath}
\usepackage{bm}
\usepackage{times}
\usepackage{color}
\usepackage{bm}
\usepackage{soul}
\usepackage{array}
\usepackage{float,placeins}
\usepackage{siunitx,marvosym}

\newcommand\Tstrut{\rule{0pt}{2.6ex}}         
\newcommand\Bstrut{\rule[-0.9ex]{0pt}{0pt}}   

\begin{document}


\title{Evaluating the Effects of Structural Disorder on the Magnetic Properties of Nd$_{2}$Zr$_{2}$O$_{7}$ }



\author{Eli Zoghlin}
\author{Julian Schmehr}
\author{Collin Holgate}
\affiliation{Materials Department, University of California, Santa Barbara, California 93106, USA}

\author{Rebecca Dally}
\affiliation{NIST Center for Neutron Research, National Institute of Standards and Technology, Gaithersburg, Maryland 20899, USA}

\author{Yaohua Liu}
\affiliation{Neutron Scattering Division, Oak Ridge National Laboratory, Oak Ridge Tennessee 37831, USA}

\author{Geneva Laurita}
\affiliation{Department of Chemistry and Biochemistry, Bates College, Lewiston, Maine 04240, USA}

\author{Stephen D. Wilson}
\email[]{stephendwilson@ucsb.edu}
\affiliation{Materials Department, University of California, Santa Barbara, California 93106, USA}



\date{\today}

\begin{abstract}
Motivated by the variation in reported lattice parameters of floating-zone-grown Nd$_{2}$Zr$_{2}$O$_{7}$ crystals, we have performed a detailed study of the relationship between synthesis environment, structural disorder, and magnetic properties. Using a combination of polycrystalline standards, electron-probe microanalysis and scattering techniques, we show that crystals grown under atmospheric conditions have a reduced lattice parameter --  relative to pristine polycrystalline powders -- due to occupation of the Nd-site by excess Zr (\textit{i.e.} ``negative'' stuffing). In contrast, crystals grown under high-pressure Ar are nearly stoichiometric with an average lattice parameter approaching the polycrystalline value. While minimal disorder of the oxygen sublattices is observed on the scale of the average structure, neutron pair-distribution function analysis indicates a highly local disorder of the oxygen coordination, which is only weakly dependent on growth environment. Most importantly, our magnetization, heat capacity and single-crystal neutron scattering data show that the magnetic properties of crystals grown under high-pressure Ar match closely with those of stoichiometric powders. Neutron scattering measurements reveal that the signature of \textit{magnetic moment fragmentation} -- the coexistence of “all-in-all-out” (AIAO) magnetic Bragg peaks and diffuse pinch-point scattering due to spin-ice correlations -- persists in these nearly stoichiometric crystals. However, in addition to an increased AIAO transition temperature, the diffuse signal is seemingly stabilized and remains nearly unchanged upon warming to 800 mK. This behavior indicates that both the AIAO magnetic order and spin-ice correlations are sensitive to deviations of the Nd stoichiometry.
\end{abstract}


\maketitle

\section{Introduction \label{intro}}
Rare-earth pyrochlore oxides (A$_{2}$B$_{2}$O$_{7}$, where A = rare-earth and B = a wide range of transition metals) have long sustained a variety of research efforts within the condensed matter community \cite{gardner2010,subramanian1983,ramirez1994,rau2019}. Recently, experimental \cite{gaudet2019, sibille2020,gao2019,mauws2018} and theoretical \cite{huang2014,li2017,yao2020} work has highlighted novel magnetic and electronic behaviors related to the dipolar-octupolar ground state doublet of certain (Kramers) rare-earth ions. Particular attention has been focused on Nd$_{2}$Zr$_{2}$O$_{7}$ following experimental observation of behavior consistent with a picture of \textit{magnetic moment fragmentation}, where different “fragments” of the Nd$^{3+}$ moment operate as separate degrees of freedom.

The idea of magnetic moment fragmentation proposed in the context of emergent Coulomb phases is based on the crystallization of monopole excitations from a spin ice ground state \cite{brooks2014}. In this theory, a Helmholtz decomposition is used to divide the magnetization field into ``divergence-full" and ``divergence-free" components. Initially, the local constraint of the spin ice state produces a magnetization that is wholly divergence-free. Fragmentation then occurs upon the crystallization of a sufficient density of monopole excitations and the development of a divergence-full contribution to the sublattice magnetization. The result is a magnetically ordered state (divergence-full) and a fluctuating Coulomb phase (divergence-free). Such a state results from a complex interplay of strong, Ising-like anisotropy, effective ferromagnetic (FM) interactions and a delicate balance of the magnetic exchange and dipolar interactions \cite{petit2016,brooks2014}.

The phenomenon of moment fragmentation can also be addressed by explicitly considering the dipolar-octupolar character of the ground state Kramers doublet of Nd$^{3+}$ ions in Nd$_{2}$Zr$_{2}$O$_{7}$ \cite{benton2016}. The differing symmetries of the resulting pseudo-spin operators ($\tau$) \cite{huang2014} result in a magnetization field derived from \textit{x} and \textit{z} terms which generate gapped magnetic fluctuations and magnetic order, respectively. Each component contributes differently to the divergence-full and divergence-free components of the total magnetization such that the fluctuations of these fields are decoupled \cite{benton2016}. 

Regardless of the precise theoretical description, Nd$_{2}$Zr$_{2}$O$_{7}$ shows the coexistence of magnetic order -- the antiferromagnetic (AFM) “all-in-all-out” (AIAO) state with reports of \textit{T$_{AFM}$} from 285 to 400 mK \cite{blote1969,xu2015,opherden2017, xu2019,lhotel2015} -- and a fluctuating Coulomb phase \cite{petit2016,xu2020}. In neutron scattering measurements, these states manifest as simultaneous magnetic Bragg peaks and an inelastic, diffuse ``pinch-point" scattering pattern, respectively \cite{benton2016,brooks2014,petit2016,xu2019,henley2010}. The availability of large, single-crystal samples of Nd$_{2}$Zr$_{2}$O$_{7}$ grown by the floating-zone (FZ) technique \cite{hatnean2015} has been instrumental to experimental investigations of the proposed magnetically fragmented state and its related magnetic properties \cite{petit2016,xu2019, xu2020, xu2018, hatnean2015_1, lhotel2015, lhotel2018}. However, truly verifying and understanding this state requires careful investigation of crystal quality, especially considering the well-known susceptibility of the pyrochlore lattice to disorder \cite{gardner2010, minervini2000, wilde1998}. 

\begin{figure}
 \includegraphics[width=8.6cm]{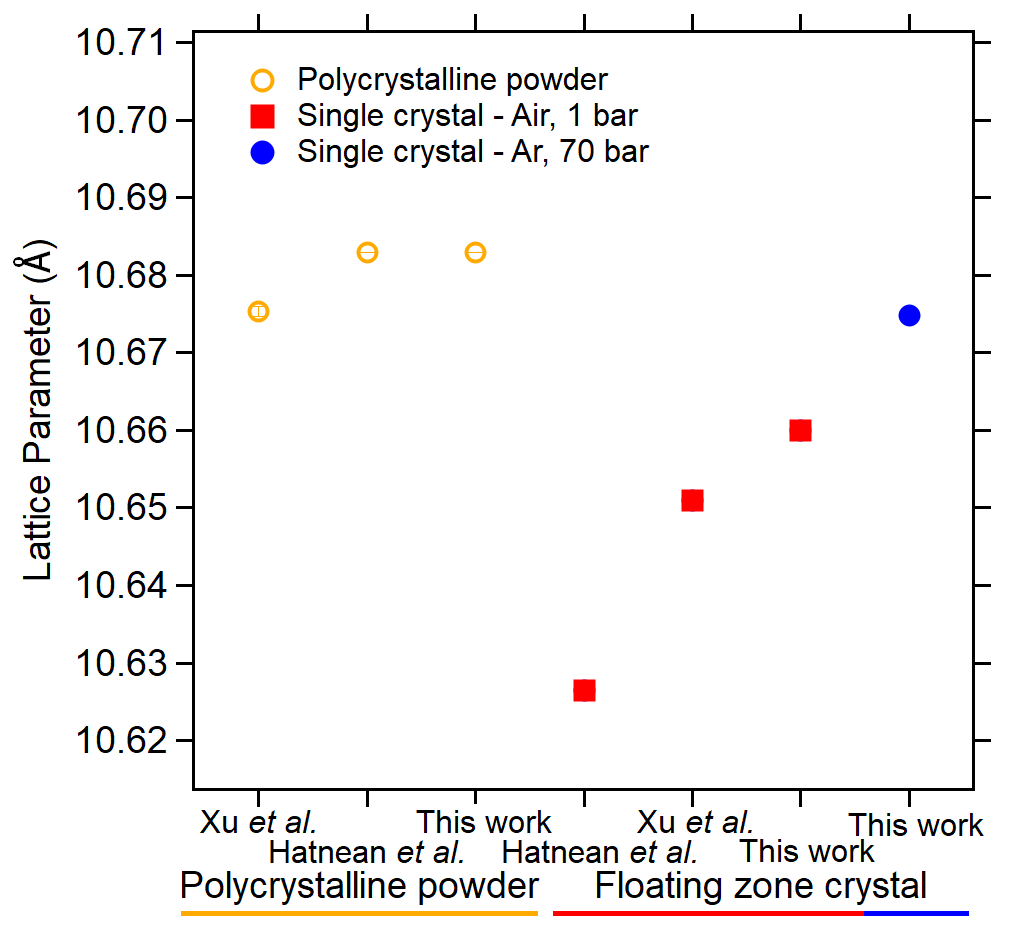}
 \caption{\label{fig:Figure1} Comparison of  lattice parameters for polycrystalline samples and single crystals from this work and the literature \cite{xu2015,hatnean2015,xu2018}. All lattice parameters were determined from laboratory XRD on powders at \textit{T} = 300 K. The error bars are contained within the symbols and represent one standard deviation}
 \end{figure}

\begin{figure}
 \includegraphics[width=8.6cm]{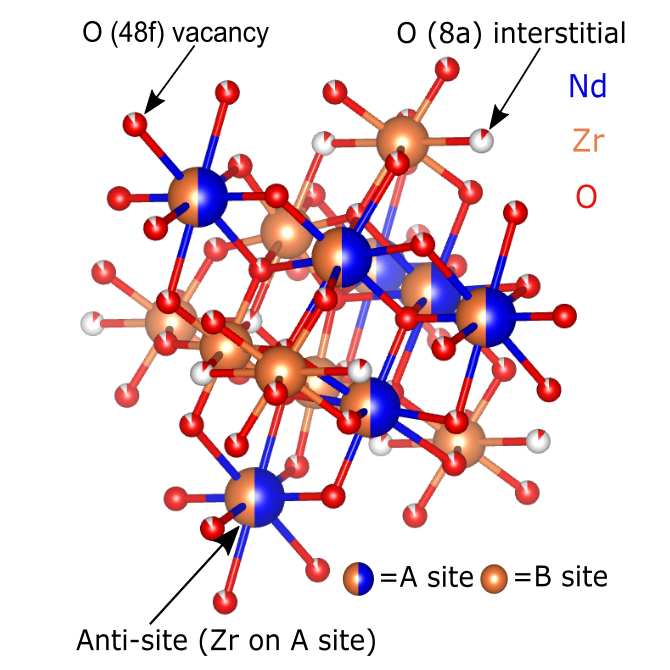}
 \caption{\label{fig:Figure2} Illustration of some of the key defect modes for Nd$_{2}$Zr$_{2}$O$_{7}$ considered in this work.}
 \end{figure}

In the case of Nd$_{2}$Zr$_{2}$O$_{7}$, deviation from unity of the Nd/Zr ratio is known to modify the cubic lattice parameter in polycrystalline samples \cite{anithakumari2016}. Reported FZ-grown crystals show a marked variation in lattice parameter \cite{xu2018,hatnean2015} relative to the value for pristine polycrystalline samples, where the Nd/Zr ratio is well-controlled and likely very close to 1 \cite{xu2015,hatnean2015} (see Fig. \ref{fig:Figure1}). FZ-grown crystals studied in the literature possess up to a $\approx$ 0.5\% smaller lattice parameter and often have substantially reduced AIAO ordering temperatures relative to powder data \cite{petit2016,lhotel2015,hatnean2015,xu2015,blote1969}. Disorder on the oxygen sublattice is also possible, where, for instance, local disorder related to the high-temperature defect-fluorite structure is known to occur \cite{payne2013, blanchard2012}. A visual summary of a number of key defect modes is shown in Fig. \ref{fig:Figure2}.

The presence of disorder can substantially renormalize interactions relative to theoretical considerations and support a complex local coexistence of states. For example, it has been suggested that intrinsic cation disorder in YbMgGaO$_{4}$ leads to the lack of observed magnetic order \cite{zhu2017}, while oxygen deficiency in Y$_{2}$Ti$_{2}$O$_{7}$ generates a paramagnetic response \cite{sala2014}. Furthermore, the FZ-growth process can lead to dramatic differences between powder and single crystal samples, such as the extensive variation (even between different FZ-grown single-crystal samples) of the heat capacity signatures in Yb$_{2}$Ti$_{2}$O$_{7}$ \cite{yaouanc2011,ross2012,bowman2019}. As such, it is important to consider what sources of structural disorder are present in Nd$_{2}$Zr$_{2}$O$_{7}$ single crystals and how this disorder might be managed by modifying the FZ-growth conditions. The ability to modify the amount of disorder leads naturally to addressing how disorder affects the bulk magnetic properties, including the experimental signatures of magnetic moment fragmentation \cite{mauws2021}.

In this paper, we leverage the high-temperature/high-pressure (HP) capabilities of laser-based FZ-growth \cite{schmehr2019} to synthesize Nd$_{2}$Zr$_{2}$O$_{7}$ samples with lattice parameters approaching those of pristine polycrystalline samples. Using a series of polycrystalline samples, we confirm the correlation between the Nd/Zr ratio and the lattice parameter and examine the effect on magnetic properties. Electron probe microanalysis (EPMA) provides direct confirmation that samples grown in atmospheric conditions (1 bar, air) with a reduced lattice parameter have Nd/Zr $<$ 1, while those grown in HP Ar (70 bar) are nearly stoichiometric. A detailed structural investigation of powdered single-crystals studied via synchrotron x-ray and neutron diffraction clarifies that this stoichiometry imbalance is accommodated by Zr occupying vacancies on the Nd sublattice. Synchrotron diffraction data also reveal a distribution of lattice parameters within FZ-grown crystals: growth in air skews the distribution towards a smaller average lattice parameter while growth in HP Ar favors a larger average lattice parameter. Although the oxygen sublattices are close to disorder free on the scale of the average structure ($<$ 2\% vacancies), pair-distribution function (PDF) analysis of the neutron data indicates 8 -- 11 mass \% of a highly local disorder of the oxygen-coordination across all samples. Crucially, we present magnetization, heat capacity, and neutron scattering measurements showing that the magnetic properties of HP Ar-grown crystals match closely with that of stoichiometric polycrystalline powders. Neutron scattering measurements on nearly stoichiometric crystals confirm the continued coexistence of AIAO magnetic Bragg peaks and a diffuse pinch-point scatterin pattern, with the latter signal persisting nearly unchanged at higher temperatures than previously reported \cite{petit2016, PhysRevLett.126.247201}. These results suggest that the stability of both the AIAO magnetic order and the spin-ice correlations are linked to Nd stoichiometry.

\section{Methods \label{methods}}

Polycrystalline samples of Nd$_{2(1-x)}$Zr$_{2(1+x)}$O$_{7}$ (\textit{x} = 0.2, 0.05, 0 and –0.1) were synthesized using solid state methods. Stoichiometric amounts of dried (900\,$^{\circ}$C, 12 hours) Nd$_{2}$O$_{3}$ (Afla Aesar, 99.99\%) and ZrO$_{2}$ (Alfa Aesar, 99.98\%) were ground together in an agate mortar and pestle and pressed into pellets at 310 MPa in a cold isostatic press. The pellets were then reacted in air in Al$_{2}$O$_{3}$ crucibles for 60 hours each at 1300, 1375 and 1450\,$^{\circ}$C with re-grinding and re-pressing between reactions. Platinum was always placed between the sample and Al$_{2}$O$_{3}$ crucible to prevent possible reaction at high temperatures. Powder X-ray diffraction (XRD) using a Panalytical Emperyan diffractometer was employed to evaluate the final product. All Rietveld refinements were performed using the TOPAS software package \cite{coelho2018}. In preparation for crystal growth, the \textit{x} = 0 powder was packed into clean rubber balloons and pressed into rods (diameter $\approx$ 4 mm, length $\approx$ 80 – 100 mm) at 310 MPa in a cold isostatic press. The rods were sintered at 1450\,$^{\circ}$C for 24 -- 48 hours, in air, on sacrificial beds of \textit{x} = 0 powder.

Single-crystal samples were then grown via the floating-zone method, using a novel, high-pressure laser furnace, the details of which are described elsewhere \cite{schmehr2019}. Growths were conducted in atmospheric conditions (1 bar, air) and at high-pressure (70 bar) of Ar (HP-Ar) or an Ar:O$_{2}$ = 80:20 mix (HP-80:20). Significant volatility was observed during the growths in air; less volatility was observed in the high-pressure growths. Other parameters were held constant across these growths, including translation rate (30 mm/hour for both feed and seed), rotation rate (counter rotation at 6 -- 8 rpm), and molten zone temperature (2400 -- 2500\,$^{\circ}$C). Note that this temperature is a relative metric using an uncorrected pyrometer. Laboratory XRD confirms that all crystals grown contained only the pyrochlore phase in the average structure. Thermally induced cracking was observed in all samples, consistent with other reports \cite{hatnean2015} and was likely exacerbated by the steep temperature gradient attained by laser heating. Varying the growth rate modified this effect, with extreme cracking and flaking at 12.5 mm/hour and very minimal cracking at 60 mm/hour, but did not otherwise modify the results with respect to the lattice parameter or phase purity. The relatively rapid growth rate represents a compromise between constraining the thermally induced cracking to only localized fractures and maintaining crystallinity.

The composition of crystals was analyzed using an EPMA (Cameca SX-100, Gennevilliers, France) equipped with wavelength-dispersive spectroscopy. Sample disks -- thin cross sections of the as-grown, cylindrical crystal boule cut perpendicular to the growth direction -- were polished with colloidal diamond to a 0.25 $\mu$m finish. The mounts were carbon coated to negate electron beam charging. A 20 kV and 20 nA beam was used for the measurements, which was defocused to 2 $\mu$m to prevent sample damage. Multiple ``spot" measurements were used to determine the composition at various points within a small (100 $\times$ 100 $\mu$m$^{2}$),  region at the center of the sample disk. Line scans along the diameter of the sample disk (from edge to edge) were used to assess any compositional variance along the boule diameter. The Zr and Nd concentration measurements were standardized using a synthetic zircon (ZrSiO$_{4}$) standard and Nd-glass standard, respectively. In the calculation of the average formula the O was assumed to be fully stoichiometric (\textit{i.e.} Nd$_{x}$Zr$_{y}$O$_{7}$). We emphasize that, due to the use of standards, the EPMA data provide an independent measurement of the Nd and Zr concentrations and the resulting formula need not show $x + y = 4$ to match the ideal formula.
 
Structural characterization was undertaken using synchrotron XRD and time-of-flight (TOF) neutron diffraction on powdered crystal boules. The synchrotron XRD measurements were performed at the 11-BM beamline of the Advanced Photon Source (APS) at \textit{T} = 100 K. To minimize absorption effects, the powder was diluted with SiO$_{2}$ at a ratio of 1:7 (molar), yielding $\mu$R $\approx$ 0.6 , before being loaded into 0.8 mm diameter Kapton tubes. TOF neutron powder diffraction data were collected using the POWGEN diffractometer at Oak Ridge National Laboratory, Spallation Neutron Source (SNS). The wavelength distribution of the incident beam was centered at 1.5 \r{A}, with a minimum of 0.967 \r{A} and maximum of 2.033 \r{A}, providing a \textit{Q} range of $\approx$ 0.43 -- 12.94 \r{A}$^{-1}$.  The powder (2.5 -- 3 g of ground FZ crystal) was sealed in 6 mm diameter V cans and measured at \textit{T} = 100 K. The POWGEN data was reduced for pair-distribution function (PDF) analysis using Mantid \cite{arnold2014}. Analysis of the total scattering data was then performed using the PDFgui software suite over various increments of a total \textit{r}-range of 1.5 -- 41.5 \r{A} \cite{farrow2007}. 

Bulk magnetic properties were measured on $\approx$ 10 mg of powder (both polycrystalline powders and ground crystals) mounted in polypropylene capsules using a MPMS3 Quantum Design SQUID magnetometer. Well-ground crystals were used to eliminate any effects of crystalline anisotropy and for comparison to the nominally stoichiometric polycrystalline sample. High-field \textit{M(H)} measurements were carried out using a vibrating sample magnetometer (VSM) mounted on a Quantum Design DynaCool Physical Properties Measurement System (PPMS) equipped with a 14 T magnet. We only make comparisons to the literature between data sets collected using the same measurement technique (SQUID or VSM). Measurements of the low temperature specific heat (\textit{C$_{P}$}) were carried out using a Quantum Design Dynacool PPMS equipped with a dilution refrigerator insert. The samples used for the measurement were taken from the center of the crystal boules with masses $<$  0.5 mg. Apiezon N-grease was used to provide thermal contact.

Single-crystal neutron diffraction was performed using the instrument CORELLI at the SNS. An intial survey of samples grown under different environments was performed at the NIST Center for Neutron Research using the cold neutron triple-axis diffractometer SPINS in order to inform the CORELLI experiment. The sample consisted of a piece extracted from the crystal boule with length $\approx$ 2.5 cm and mass $\approx$ 1.3 g. Copper foil was used to attach the crystal to an oxygen-free copper mount to ensure adequate thermalization; in the CORELLI experiment a Cd mask was used to reduce scattering from the mount and N-grease was also placed at the interface between the crystal and the mount. The crystal was then mounted in a liquid helium cryostat with a dilution insert, providing a base temperature of $\approx$ 38 mK. Data across a large area of reciprocal space were collected at \textit{T} = 38, 500, and 800 mK by rotating the crystal around the [$1\bar{1}$0] direction ($\omega$ rotation) in 1$^{\circ}$ steps with a count time of $\approx$ 3 minutes/step. A magnetic order parameter on the (220) reflection was collected by fixing $\omega$ such that the peak could be measured. The CORELLI data was reduced using Mantid \cite{taylor2012}.  Confidence intervals for all error bars and uncertainties represent plus and minus one standard deviation. 

\section{Experimental Results \label{results}}

\subsection{Structural Properties Characterization \label{structure}}

\begin{figure}
 \includegraphics[width=8.6cm]{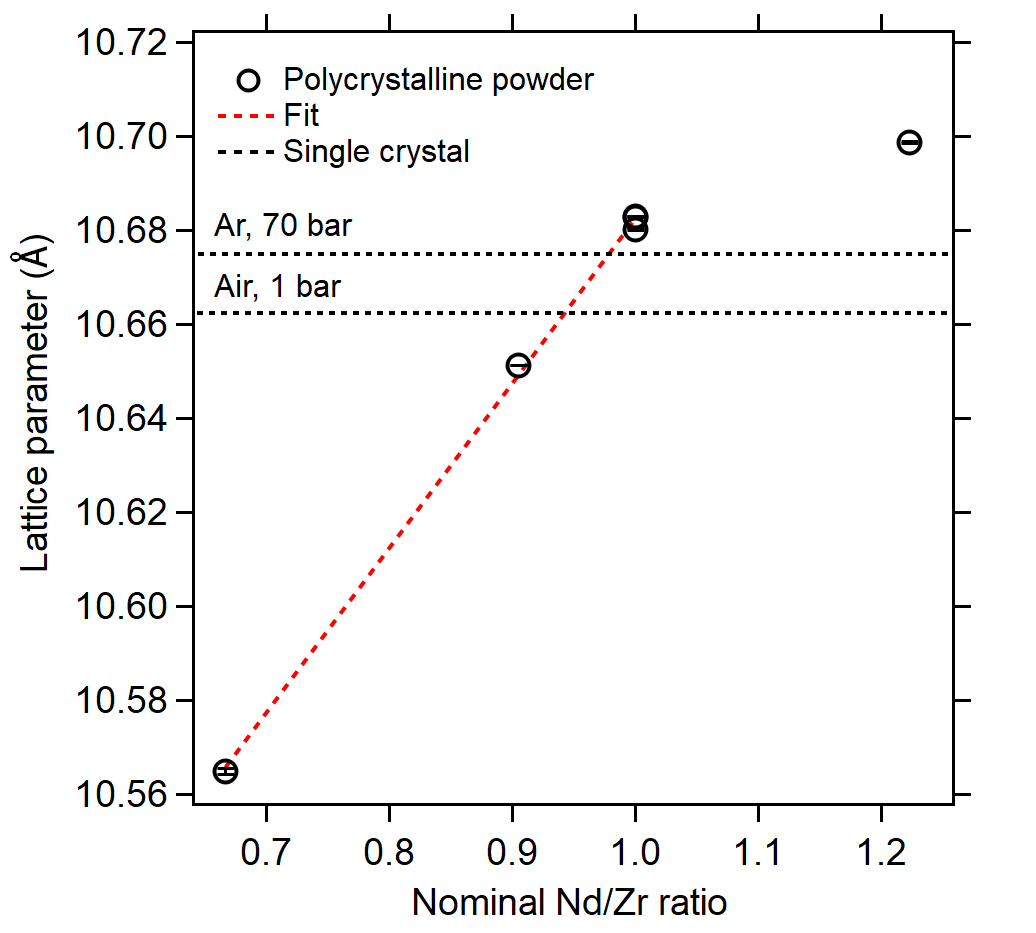}
 \caption{\label{fig:Figure3} Refined room temperature lattice parameters from laboratory XRD versus nominal Nd/Zr ratio for polycrystalline samples of Nd$_{2(1-x)}$Zr$_{2(1+x)}$O$_{7}$ with \textit{x} =  0.2, 0.05, 0, and -0.1. The error bars are contained within the symbols and represent one standard deviation. The Nd-enriched polycrystalline sample is biphasic (Nd$_{2}$O$_{3}$ + ``Nd$_{2.2}$Zr$_{1.8}$O$_{7}$") as is expected from the change in slope. The dashed red line is a linear fit to the phase pure samples while the dashed black lines indicate the lattice parameters found by laboratory XRD for single crystal samples grown in this work.}
 \end{figure}

To provide a baseline in understanding how the Nd/Zr ratio impacts the properties of Nd$_{2(1-x)}$Zr$_{2(1+x)}$O$_{7}$, a series of polycrystalline samples (``standards") were first prepared. The lattice parameter for the Nd-deficient (\textit{x} $>$ 0) samples decreases linearly with decreasing Nd/Zr relative to the \textit{x} = 0 samples (Fig. \ref{fig:Figure3}). For the Nd-enriched sample (\textit{x} $<$ 0), the lattice parameter deviates from the linear trend, likely reflecting the fact that it was biphasic, with $\approx$ 5 mass \% Nd$_{2}$O$_{3}$. However, it is still clearly enhanced relative to the stoichiometric sample. Due to the lower processing temperature, the Nd/Zr ratios are assumed as the nominal values of the polycrystalline samples and these values were fixed during the refinement of the powder data used to determine the lattice parameters.  The results are consistent with those of a previous study on Nd$_{2(1-x)}$Zr$_{2(1+x)}$O$_{7}$ powders \cite{anithakumari2016}. 

A comparison can then be performed between the composition and structure of single crystals grown in a variety of environments.  The lattice parameters of single crystals (all based on laboratory XRD) reported in the literature and in this work are summarized in Fig. \ref{fig:Figure1}, with polycrystalline values plotted for comparison. Reported values for low-pressure, air-grown samples vary considerably, ranging from 10.627 \cite{hatnean2015_1} to 10.662 \r{A} (this work), but are in all cases reduced compared to the polycrystalline samples ($\approx$ 10.68 \r{A}). In contrast, samples grown in 70 bar of Ar show an elevated lattice parameter (10.675 \r{A}) which approaches that of the polycrystalline samples. 

\begin{figure}
 \includegraphics[width=8.6cm]{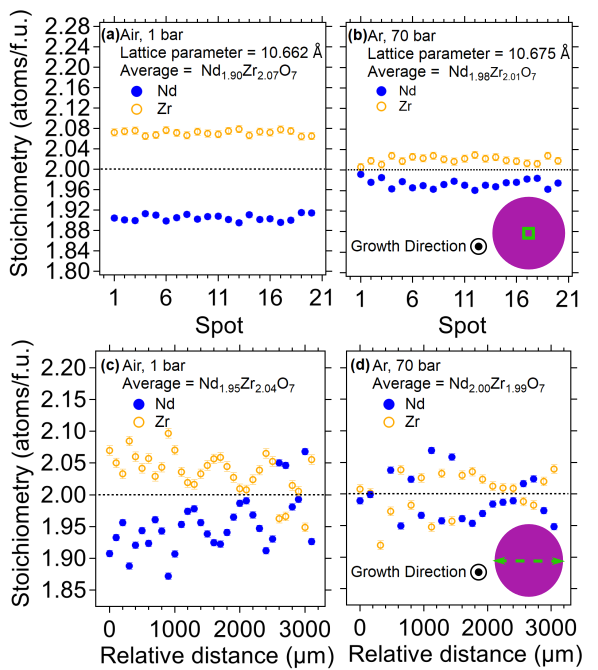}
 \caption{\label{fig:Figure4} EPMA data on sample disks of the FZ crystals cut perpendicular to the growth direction. Error bars represent the statistical error reported from the measurement software. The top panels show results for a number of "spot" measurements taken at different points within a small central region of the sample disk for the (a) air and (b) HP-Ar samples. The measurement region is represented by the green square (not to scale) in the inset of (b). The bottom panels show results from line scans for the (c) air and (d) HP-Ar samples. The direction of the scan is shown by the green arrow in the inset of (d). The distance is measured relative to one edge of the sample disk, extending across the diameter of the slice.
 	}
 \end{figure}

Nd and Zr stoichiometries for two single-crystals (air and HP-Ar), determined by the EPMA measurements, are shown in Fig. \ref{fig:Figure4}. Spot measurements, taken in a small, central area of the boule, show a consistent difference in stoichiometry between the two samples. The air-grown sample has Nd/Zr = .918 (average formula = Nd$_{1.90}$Zr$_{2.07}$O$_{7}$) while the HP Ar-grown sample has Nd/Zr = .985 (average formula = Nd$_{1.98}$Zr$_{2.01}$O$_{7}$).  In contrast, line scans, taken across a larger extent of the sample, show an increased range of values relative to the spot measurements. The spread of the Nd and Zr stoichiometric coefficients is roughly 0.15 for the air-grown sample and 0.10 for the HP-Ar sample. However, the average stoichiometry of the line scans mirrors the behavior seen in the spot measurements: the air-grown sample has Nd/Zr $<$ 1 while the HP-Ar sample has, on average, Nd/Zr closer to 1.

\begin{figure*}
 \includegraphics[width=\textwidth]{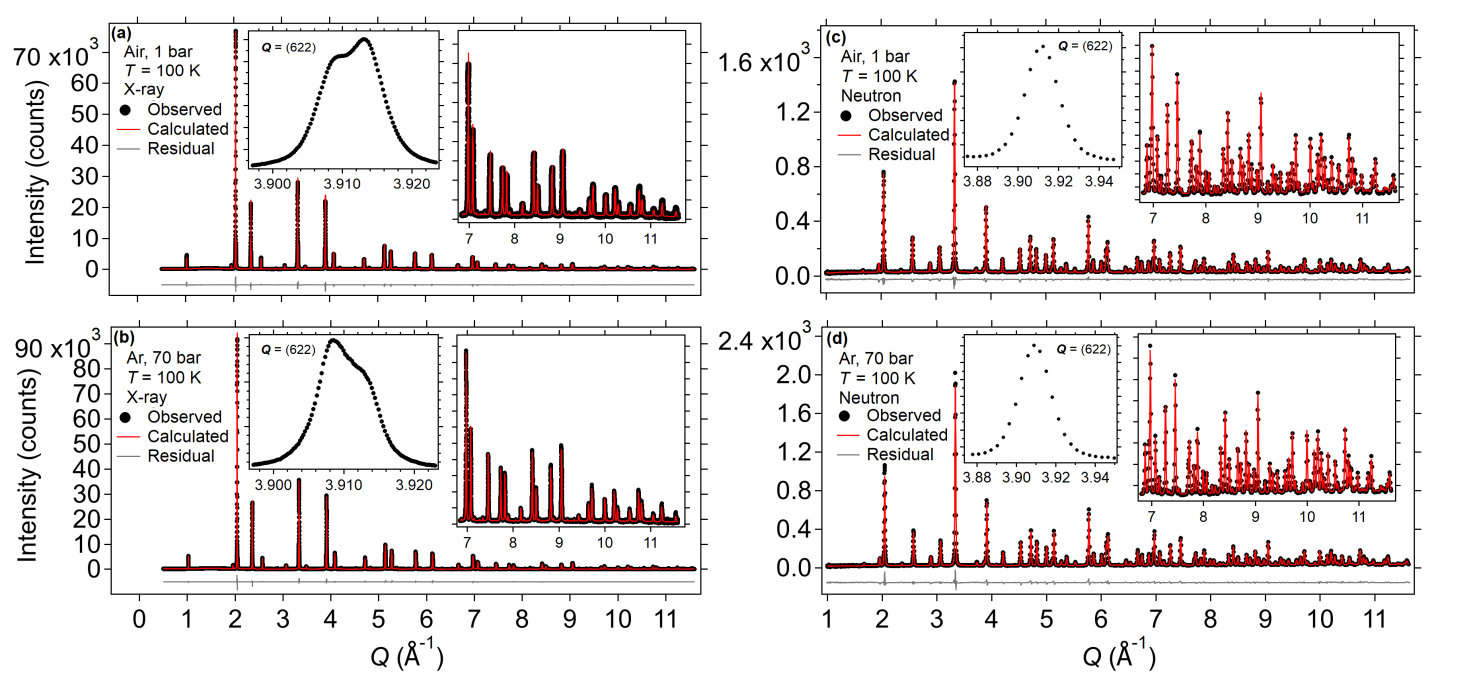}
 \caption{\label{fig:Figure5} Rietveld refinement of synchrotron XRD data collected from powdered FZ-boules grown under (a) air and (b) HP-Ar. Refinement of TOF neutron diffraction data for powders of the same boules, (c) air and (d) HP-Ar. The left inset shows a close view of the (622) peak, illustrating the the variable peak shapes observed in the XRD data, but not the TOF neutron data. The right inset shows a closer view of the fit to the high-\textit{Q} data.}
 \end{figure*}

The synchrotron X-ray and TOF neutron powder diffraction data for the HP-Ar and air-grown samples were analyzed with a simultaneous Rietveld refinement to provide deeper insight into structural disorder on both the cation (Nd, Zr) and anion (O) sites. The refined patterns resulting from the structural model described below are shown in Fig. \ref{fig:Figure5}, with the results for the refined parameters summarized in Table \ref{tab:Table1}. For both samples, all peaks could be matched to the pyrochlore structure (Fd$\bar{3}$m, space group number 227). In this structure the Nd and Zr cations each occupy a distinct, high-symmetry, site forming separate corner-sharing sublattices of Nd and Zr tetrahedra. The O anion occupies two sites, referred to here as O1 (Wyckoff position 48\textit{f}) and O2 (Wyckoff position 8\textit{b}). The pyrochlore unit cell is predominately constrained by symmetry and provides only \textit{u}, the \textit{x}-coordinate of the O1-site and the lattice parameter (\textit{a}) as free parameters.

Establishing a structural model for the co-refinement was complicated by the observation of a split-peak structure in the higher momentum resolution X-ray data, but not the neutron data, of both samples. To model this feature, the data were refined using two pyrochlore phases with different lattice parameters -- a similar approach has been taken with other site-mixed pyrochlores \textit{e.g.} Ho$_{2(1+x)}$Ti$_{2(1-x)}$O$_{7}$ \cite{baroudi2015}. We note here that comparable high-resolution X-ray data for the HP-80:20 sample were also collected and were not well fit by two pyrochlore phases, requiring the addition of further pyrochlore phases to improve the fit. In the interest of establishing a parsimonious and consistent structural model for the co-refinement we restrict our comparison to the air and HP-Ar-grown samples here. 

Since the TOF neutron data show only one peak, due to the coarser resolution, the lattice parameters were wholly determined by the X-ray data and fed into the calculated pattern for the neutron data. A somewhat analogous approach was taken with respect to the calculated phase fractions of each pyrochlore phase. The X-ray scale factors for the two phases (\textit{S$_{X1}$} and \textit{S$_{X2}$}) were refined independently while only one neutron scale factor (\textit{S$_{N1}$}, corresponding to the majority phase, as determined by the X-ray data) was refined. The other neutron scale factor (\textit{S$_{N2}$}) was then calculated as $S_{N2} = (S_{X2}\, /\, S_{X1})$$\,\times\, S_{N1}$ to ensure the same phase fractions as determined in the X-ray data. Other values, including the site occupancies, anisotropic displacement parameters (ADPs), and \textit{u} (O1 \textit{x}-position) were refined globally (\textit{i.e.} against both patterns). Based on the EPMA data, the Nd A-site occupancy was fixed to the average value of the spot measurements (Fig. \ref{fig:Figure4}) while the Zr B-site occupancy was fixed to 1. Occupation of the A-site by Zr was then allowed later in the analysis, constrained by a maximum total occupancy of 1. Refinement of the O1 and O2 site occupancies showed the O2 site to be fully occupied while the O1 site has a small number of vacancies ($<$2 \%), consistent with the general trend in pyrochlores of oxygen vacancies occurring preferentially on the 48\textit{f}-site \cite{payne2013,ubic2008,wilde1998}. The two pyrochlore phases were constrained to have the same set of ADPs and the same value of \textit{u}. 

\begin{figure}
 \includegraphics[width=8.6cm]{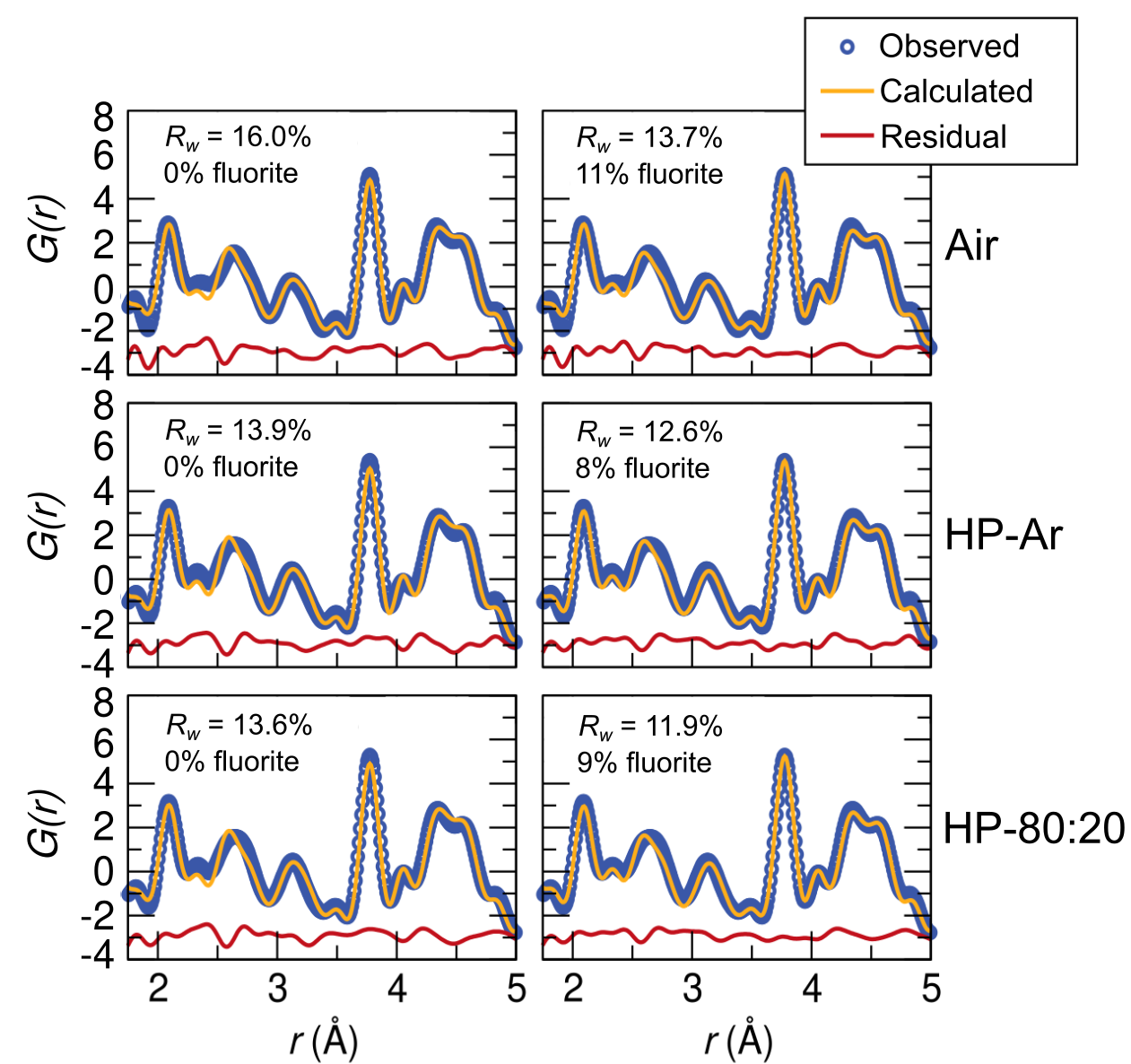}
 \caption{\label{fig:Figure6} Fits to the neutron PDF spectra in the \textit{r} = 1.5 -- 5 \r{A} range for the air (top), HP-Ar (middle) and HP-80:20 (bottom) samples. Defect fluorite phase fraction is quoted in mass \%.}
 \end{figure}

To gain insight into \textit{r}-dependent behavior (\textit{i.e.} local disorder that does not appear in the average structure) PDF spectra were refined using a “box-car” scheme. Fits to the \textit{G(r}) data were performed within a set \textit{r}-range (\textit{r$_{min}$} -- \textit{r$_{max}$}  = 5 \r{A}) at various length scales (e.g. 0 -- 5 \r{A}, 5 -- 10 \r{A}). Refinement using the ideal pyrochlore structure results in relatively flat difference curves and reasonable fit qualities (\textit{R$_{W}$} $\approx$ 10\%) for most of the \textit{r}-range (not shown). However, a notable decrease in fit quality was found for all samples within the \textit{r} = 1.5 -- 6.5 \r{A} range (\textit{R$_{W}$} $\approx$ 13 -- 16\%), driven mainly by a poor fit to the peak at \textit{r} = 2.6 \r{A}.  Calculation of site-specific partial PDF contributions reveal this peak to be composed of both Nd -- O and O -- O correlations, with the O -- O correlations being purely due to correlations between O1 sites. Refining the O1 site occupation showed a minor reduction in O1 occupancy ($<$ 3\%) with only marginal improvement to \textit{R$_{W}$}.  However, inclusion of a secondary defect fluorite phase for the fit in the \textit{r} = 1.5 -- 5 \r{A} range resulted in a decrease in \textit{R$_{W}$} and an observable improvement in the fit of the peak at 2.6 \r{A} (Fig. \ref{fig:Figure6}). Refinement of the amount of the defect fluorite structure shows the highest phase fraction (mass \%) for the air-grown sample (11.2 $\pm$ 0.22 \%), the lowest for the HP-Ar sample (8.2 $\pm$ 0.18 \%) and an intermediate amount (8.8 $\pm$ 0.18 \%) for the HP-80:20 sample. Inclusion of this phase in the highest \textit{r}-range (36.5 -- 41.5 \r{A}) produced only a marginal improvement to the fit (change in \textit{R$_{W}$} of $<$ 1). An increase in the amount of defect fluorite is also accompanied by additional positional disorder on the O1 site, indicated by an increase in the isotropic displacement parameter. The implications of this model are further addressed in Section \ref{disc} of this paper.

\subsection{Magnetic Properties Characterization \label{magnetic}}

\begin{figure}
\includegraphics[width=8.45cm]{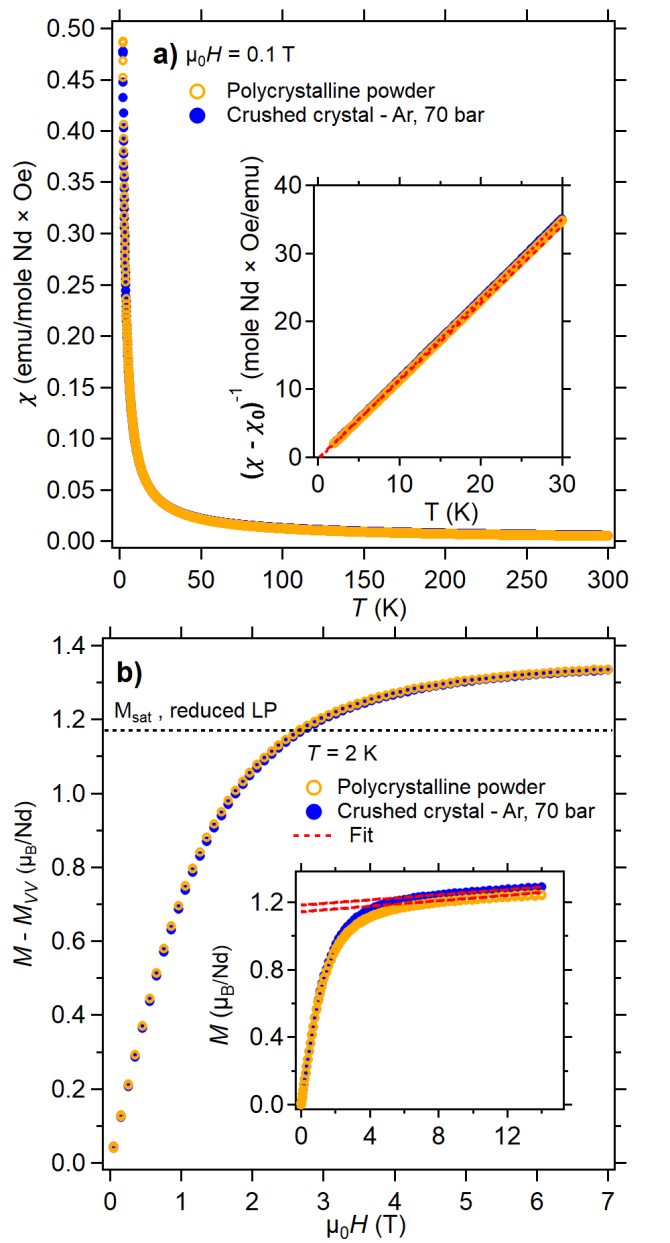}
\caption{\label{fig:Figure7} (a) Magnetic susceptibility (1 emu = 10$^{-3}$A$\,$m$^{2}$) for powder of the HP Ar-grown crystal and the \textit{x} = 0 polycrystalline powder taken at $\mu_{0}H$ = .1 T (1 T = 10,000 Oe). The inset shows the inverse susceptibility for the same; the dashed red lines are the Curie-Weiss fits. (b) Magnetization verus field data for powder of the HP Ar-grown sample and the polycrystalline sample. The Van Vleck contribution has been subtracted. The dashed black line is the expected powder saturation value for a previously reported reduced lattice parameter (LP) single-crystal sample \cite{hatnean2015_1}. The inset shows magnetization data out to  $\mu_{0}H$ = 14 T prior to subtraction of the Van Vleck contribution. The dashed red line is the linear fit from 8 -- 14 T, which has been extended to low-field.}
 \end{figure}

Characterization of the bulk magnetic properties of the HP-Ar crystal and \textit{x} = 0 powder are shown in Fig. \ref{fig:Figure7}. The two samples show nearly identical $\chi(T)$ data. A Curie-Weiss analysis was performed by first fitting $\chi(T)$ to the equation $\chi_{VV} + C\,/\,(T\, -\, \Theta_{CW})$, with $\chi_{VV}$ capturing the temperature-independent contribution of the Van Vleck paramagnetism. $\chi(T) - \chi_{VV}$ was then fit to $C\,/\,(T\, -\, \Theta_{CW})$ in the range 10 $\leq$ \textit{T} $\leq$ 30 K to avoid the influence of short-range magnetic correlations and crystal field effects \cite{xu2015}. This gave $\Theta_{CW}$  = 0.309(5) K and $\mu_{eff}$ = 2.60(2) $\mu_{B}$/Nd for the HP-Ar crystal and $\Theta_{CW}$   = 0.386(4) K, $\mu_{eff}$ = 2.61(3) $\mu_{B}$/Nd for the powder. 

 \begin{figure}
\includegraphics[width=8.6cm]{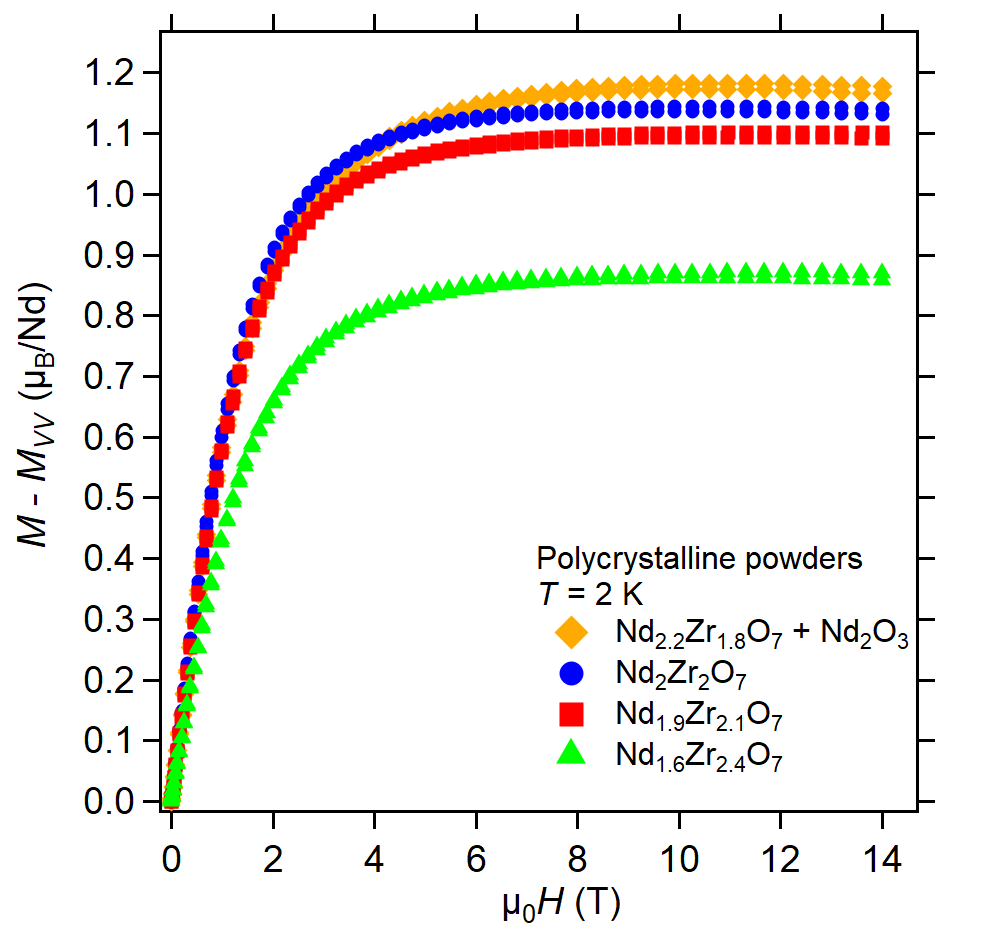}
\caption{\label{fig:Figure8} Magnetization versus field (VSM) measurements to $\mu_{0}H$ = 14 T at \textit{T} = 2 K on polycrystalline Nd$_{2(1-x)}$Zr$_{2(1+x)}$O$_{7}$ samples. The Van Vleck contribution ($M_{VV}$) has been subtracted.}
\end{figure}

Isothermal $M(\mu_{0}H)$ data at 2 K for the powder standards are plotted in Fig. \ref{fig:Figure8}. The Van Vleck contribution of the Nd$^{3+}$ ion has been subtracted using the slope of a linear fit made in the range 8 -- 14 T, beyond the field values where the Nd moments saturate. The resulting saturation magnetization, $M_{sat}$, is lower for the Nd-deficient samples, relative to the \textit{x} = 0 sample, and higher for the Nd-enriched sample. Curie-Weiss analysis (not shown) indicates that $\mu_{eff}$ is also decreased for Nd-deficient samples. The data for the HP-Ar crystal and \textit{x} = 0 powder are shown together in the main panel of  Fig. \ref{fig:Figure7}(b). The two are, again, very similar and both yield a $M_{sat}$($\mu_{0}H$ = 7 T) = 1.33 $\mu_{B}$/Nd. For further comparison, a value for $M_{sat}$($\mu_{0}H$ = 5 T) = 1.27 $\mu_{B}$/Nd has been reported previously for polycrystalline Nd$_{2}$Zr$_{2}$O$_{7}$ \cite{xu2015}, which agrees reasonably well with our data at the same field value. The black, dashed line is the approximate, powder-averaged, $M_{sat}$($\mu_{0}H$ = 7 T) for an air-grown single-crystal sample with a particularly low lattice parameter (10.627 \r{A}) \cite{hatnean2015_1}. This estimate is calculated based on the reported values of $M_{sat}$($\mu_{0}H$ = 7 T) along the [111], [110] and [100] directions: $M_{sat} = (6M_{[100]}\, +\,12M_{[110]}\, +\,8M_{[111]})\,⁄\,26$ = 1.19 $\mu_{B}$/Nd. The same calculation for a different single crystal sample grown in air (lattice parameter = 10.651 \r{A}) also shows a value less than that of the HP-Ar crystal \cite{xu2019}. 

\begin{figure}
\includegraphics[width=8.6cm]{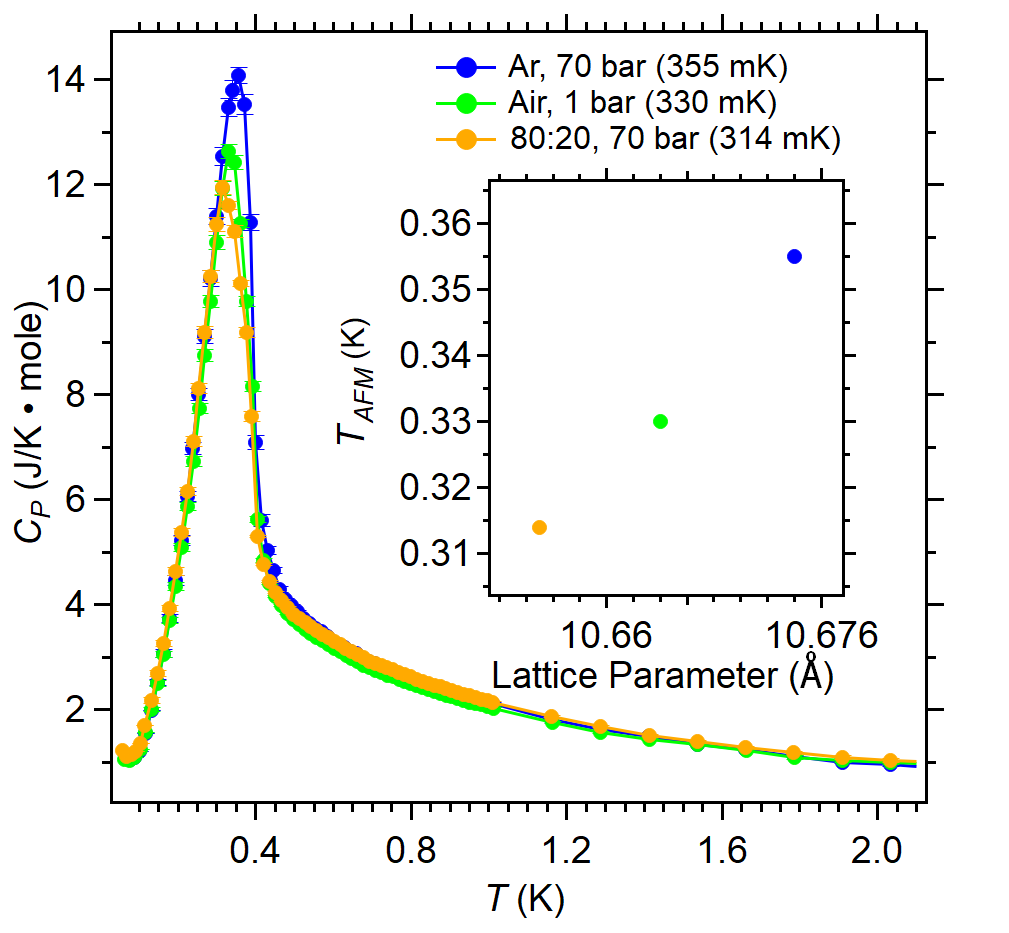}
\caption{\label{fig:Figure9} Low-temperature $C_{P}$\textit{(T)} data on the air, HP-Ar, and HP-80:20 samples. The inset shows $T_{AFM}$  (taken as the peak of the $C_{P}$\textit{(T)} curve) versus lattice parameter from laboratory XRD.}
 \end{figure}

Extending comparisons of magnetic properties to lower temperature and into the ordered state, low-temperature $C_{P}$\textit{(T)} data were collected for air, HP-Ar and HP-80:20 samples (Fig. \ref{fig:Figure9}). $T_{AFM}$, the onset of long-range magnetic order, is identified as the peak in $C_{P}$\textit{(T)}, and the data across the series of crystals show that $T_{AFM}$ is lowest for the crystal grown under HP-80:20 (314 mK) and highest for the sample grown under HP-Ar (355 mK). Compared to the air-grown crystal, the magnetic entropy associated with the phase transition ($\Delta$$S$, integrated up to 2 K) is increased by $\approx$ 6\% for the HP-Ar crystal. This reflects the increase in the ordered moment due to the reduction in off-stoichiometry.

\begin{figure}
\includegraphics[width=8.6cm]{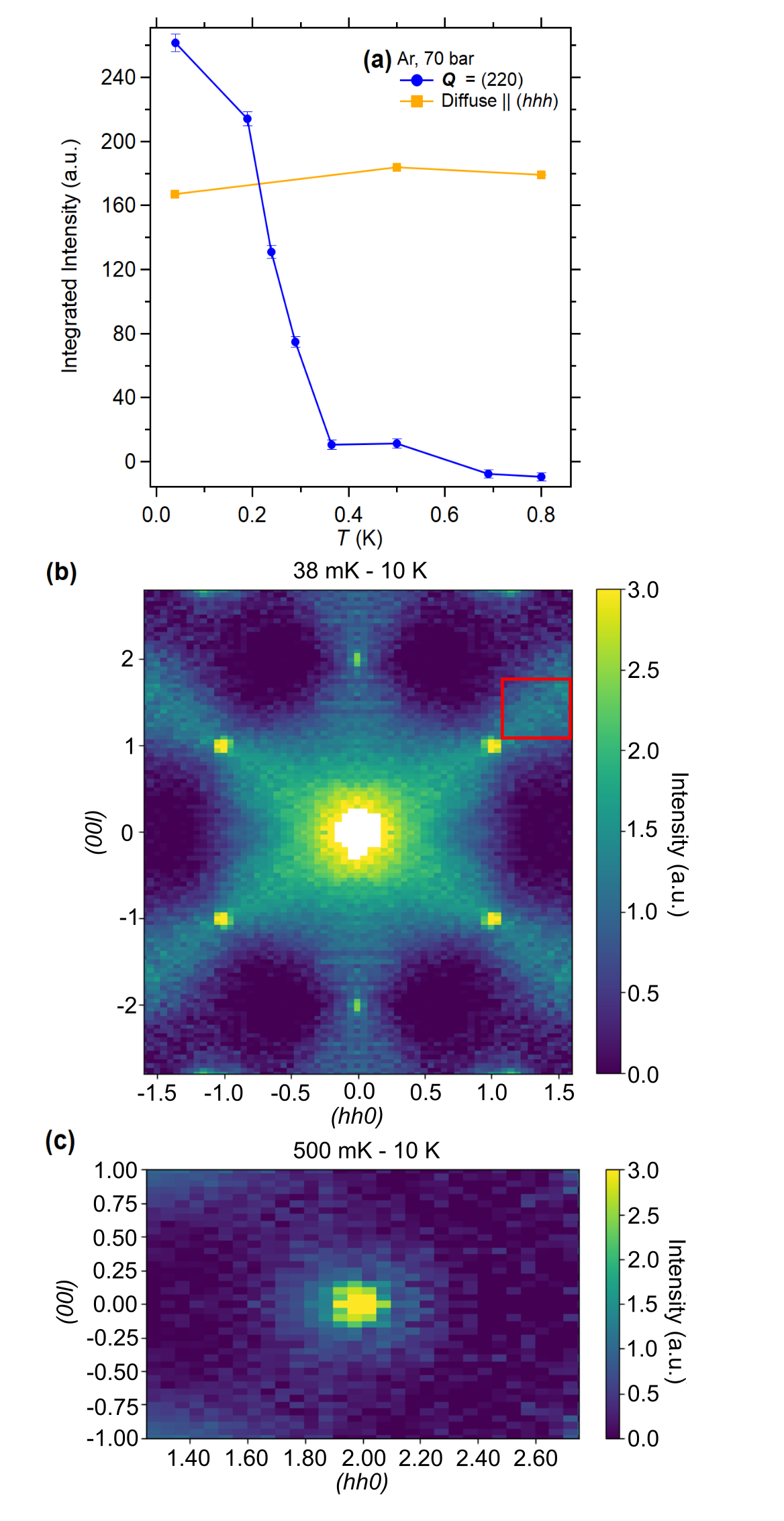}
\caption{\label{fig:Figure10} (a) Order parameters collected for a HP-Ar single-crystal. The blue markers show the evolution of the integrated intensity at the (220) reflection. The error bars are extracted from the uncertainty in the Gaussian fit parameters. The orange markers show the same for the diffuse scattering along the (\textit{hhh}) direction (excluding Bragg peaks and scaled to be comparable) (b) Map of the static diffuse scattering in the (\textit{hhl}) scattering plane at 38 mK. The red box indicates the integration area. A 10 K background scan has been subtracted. The concentrated intensity at the (111) is due to the imperfect removal of the nuclear Bragg scattering. (c)  Map of of the diffuse scattering around the (220) at 500 mK. Both (b) and (c) have been averaged according to the symmetry of the lattice. }
 \end{figure}

Further characterization of the magnetic order within the HP-Ar sample was accomplished by a single-crystal neutron scattering measurement at CORELLI. The integrated intensities of the (220) reflection are plotted in Fig. \ref{fig:Figure10}(a). The integrated intensity increases markedly below \textit{T} = 365 mK. Full $\omega$-scans, covering a much larger area of the (\textit{hhl}) scattering plane, also show a clear increase in intensity of the (113) reflections between \textit{T} = 500 mK and 38 mK. Diffuse scattering in the (\textit{hhl}) plane at \textit{T} = 38 mK is shown in Fig. \ref{fig:Figure10}(b) with a high-temperature data set (\textit{T} = 10 K) subtracted. The data have been reduced using the cross-correlation analysis associated with the statistical chopper in place at CORELLI \cite{ye2018}. While the diffuse scattering is naively not polluted by thermal diffuse scattering, the energy of the gapped pinch-point mode near 50 $\mu$eV \cite{petit2016,xu2019} is still integrated over in the finite energy window of the cross-correlation analysis. The pattern is qualitatively very similar to previous reports, showing pinch-point scattering characteristic of the spin-ice correlations within the fluctuating Coulomb phase \cite{petit2016,xu2019,xu2020}. 

To quantify the evolution of the pinch-point pattern with temperature, the diffuse scattering was integrated in a limited rectangular region containing the arm of diffuse scattering along the (\textit{hhh}) direction; the result is shown in Fig. \ref{fig:Figure10}(a). Separate from this pinch-point scattering, a second region of relatively isotropic, diffuse signal was also observed at \textit{T} = 500 mK (above $T_{AFM}$) around the (220) and (113) reflections (Fig. \ref{fig:Figure10}(c)). Further above $T_{AFM}$  (\textit{T} = 800 mK), this diffuse signal broadens considerably and begins to bleed into the persistent pinch-point pattern. The two diffuse scattering signals above $T_{AFM}$ suggest separate regimes of short-range magnetic correlations associated with the AIAO and spin-ice order parameters, which we discuss in the next section.

\section{Discussion \label{disc}}

The significant range of reported lattice parameters for FZ-grown Nd$_{2}$Zr$_{2}$O$_{7}$ single crystals (Fig. \ref{fig:Figure1}) is an indicator of variable, synthesis-dependent, disorder. Characterization of Nd$_{2(1-x)}$Zr$_{2(1+x)}$O$_{7}$ polycrystalline samples, which have more well-controlled stoichiometries than the FZ-grown samples, provides a benchmark if this disorder occurs on the Nd and Zr sublattices. Considering the results from these samples, varying cation stoichiometry is the most likely explanation for the range of reported lattice parameters of FZ-grown samples. The melting point of Nd$_{2}$Zr$_{2}$O$_{7}$ ($\approx$ 2300 -- 2350 \,$^{\circ}$C \cite{hatnean2016}) requires a high molten zone temperature leading to significant oxide vapor pressure during FZ-growth. Laboratory XRD on volatilized powder, collected following growth under ambient conditions, showed it to be primarily Nd$_{2}$O$_{3}$. Therefore, this volatilization preferentially depletes the molten zone of Nd$_{2}$O$_{3}$, which leads to a decrease in Nd/Zr of the grown crystal and a reduced lattice parameter. Our EPMA data confirms this picture, directly linking the reduced lattice parameter to Nd/Zr $<$ 1. In comparison, application of high-pressure Ar clearly helps to mitigate Nd$_{2}$O$_{3}$ volatility, resulting in a crystal with a larger lattice parameter which the EPMA data show to have a Nd/Zr ratio close to 1. 

As a further check, the Nd/Zr ratios calculated using the fit to lattice parameters of polycrystalline standards (Fig. \ref{fig:Figure3}) is found to compare well to the value extracted from EPMA data. The fit gives a Nd/Zr ratio of 0.94 for the air-grown sample and 0.98 for the HP-Ar sample while the EPMA average shows 0.92 and 0.99, respectively. For comparison, the fit implies Nd/Zr ratios of 0.84 and 0.91 for previously reported air-grown crystals with lattice parameters of 10.627 \r{A} \cite{petit2016,hatnean2015_1} and 10.651 \r{A} \cite{xu2019,xu2018}, respectively.  We note that these previously reported values may be anomalously low due to variance in the Nd/Zr stoichiometry across a crystal boule, particularly if only a small portion of the crystals were analyzed.  It is possible that the average lattice parameters are closer to the value of 10.662 \r{A} observed in our air grown crystal. 

 \begin{table*}[]
 \caption{\label{tab:Table1}Results from the simultaneous Rietveld refinement of the synchrotron X-ray and TOF neutron diffraction data for the HP-Ar and air-grown samples. Numbers in parenthesis represent plus and minus one standard deviation as reported from the refinement software. The data, by row, are as follows: \textit{T} = measurement temperature, $a_{H/L}$ = lattice parameters of the pyrochlore phases (\textit{L} = low lattice parameter phase, \textit{H} = high lattice parameter phase), $W_{H/L}$ = phase fraction (mass \%) of the pyrochlore phases, \textit{u} =  \textit{x}-position of the O1 (48\textit{f}) oxygen site (both phases), occupancy values for the different crystallographic sites, ADP = anisotropic atomic displacement parameters, $U_{ij}$ (both phases) for each site, $R_{WP}$ and $\chi^{2}$ = goodness of fit parameters for the individual patterns and overall refinement.}
 \begin{ruledtabular}
\begin{tabular*}{17cm}{c|cccc|cccc} 
Sample & \multicolumn{4}{c|}{Ar, 70 bar} &\multicolumn{4}{c}{Air, 1 bar} \Tstrut \\ \hline 
\textit{T (K)} & \multicolumn{4}{c|}{100} & \multicolumn{4}{c}{100} \Tstrut \\
$a_{H}$ (\r{A}) & \multicolumn{4}{c|}{10.66501(1)} & \multicolumn{4}{c}{10.66432(1)} \\ 
$W_{H}$(\%) & \multicolumn{4}{c|}{63.7} & \multicolumn{4}{c}{41.2}  \\
$a_{L}$ (\r{A}) & \multicolumn{4}{c|}{10.65254(1)} & \multicolumn{4}{c}{10.65069(1)}\\
$W_{L}$ (\%) & \multicolumn{4}{c|}{36.3} & \multicolumn{4}{c}{58.8} \\
$u$ (O1) & \multicolumn{4}{c|}{0.33542(2)} & \multicolumn{4}{c}{0.33542(2)} \\
A-site occ. & \multicolumn{2}{c}{Nd = 0.99} & \multicolumn{2}{c|}{Zr = 0.010(1)} & \multicolumn{2}{c}{Nd = 0.95} & \multicolumn{2}{c}{Zr = 0.050(1)} \\
B-site occ. & \multicolumn{4}{c|}{ Zr = 1} &  \multicolumn{4}{c}{Zr = 1} \\
O1-site occ. & \multicolumn{4}{c|}{0.997(1)} &  \multicolumn{4}{c}{0.980(1)} \\
O2-site occ.  & \multicolumn{4}{c|}{1.000(2)} &  \multicolumn{4}{c}{0.999(2)} \Bstrut \\  \hline
ADP $10^{-3}$ (\r{A}$^{2}$) & $U_{11}$ & $U_{22} = U_{33}$  & $U_{12} = U_{13}$ & $U_{23}$ & $U_{11}$ & $U_{22} = U_{33}$  & $U_{12} = U_{13}$ & $U_{23}$ \Tstrut \\ \hline
A-site &  3.97(2) & 3.97(2)  & -0.98(4)  & -0.98(4) & 5.61(2) & 5.61(2)  & -1.13(5) & -1.13(5) \Tstrut  \\
B-site &  4.45(3) & 4.45(3)  & 0.60(5)   &  0.60(5) & 4.83(2)  & 4.83(2) & 0.72(5)     & 0.72(5) \\
O1-site & 9.60(6)& 6.10(5)  &  0        &  0.63(7) & 10.37(7) & 6.80(4) & 0            & 0.63(7) \\
O2-site & 5.29(8) &  5.29(8)& 0 	       &    0      &  6.13(7) & 6.13(7)  &0             &0 \\ \hline
\multicolumn{1}{c}{} &\multicolumn{4}{c}{Goodness-of-fit parameters} & \multicolumn{4}{c}{Goodness-of-fit parameters}\Tstrut \\ \hline
& Neutron & X-ray & \multicolumn{2}{c|}{Overall} &Neutron & X-ray & \multicolumn{2}{c}{Overall} \Tstrut \\ \hline
$R_{WP}$ & 5.44 & 6.84 & \multicolumn{2}{c|}{6.15} & 4.71 & 7.91 & \multicolumn{2}{c}{6.39} \Tstrut \\
$\chi^{2}$ & 4.31 & 1.31 & \multicolumn{2}{c|}{1.66} & 3.80 & 1.50 &  \multicolumn{2}{c}{1.72} \\
 \end{tabular*}
 \end{ruledtabular}
 \end{table*}

The X-ray/neutron co-refinement results (Table \ref{tab:Table1}) provide clarification about the nature of this cation site disorder. When Zr is allowed to refine on the A-site, the refined value in both samples reached the maximum value of 1 minus the fixed Nd occupancy. In the EPMA data, we see that Zr-enrichment occurs in tandem with the Nd-deficiency: the corefinement shows that this Zr excess is accommodated by occupation of the depleted Nd sublattice by Zr. We note that in the pyrochlore literature the reverse of this defect (occupation of the B-site by the A-site ion) is referred to as “stuffing” and has been studied extensively (see, for example, \cite{baroudi2015,ross2012,bowman2019}). The situation here, termed “negative” stuffing, has been the subject of considerably fewer studies \cite{telang2018,ghasemi2018,mostaed2018}. 

The results from the two lattice parameter model are also consistent with the relationship between Nd/Zr ratio and lattice parameter derived from laboratory XRD. The HP-Ar sample, where EPMA indicates Nd/Zr close to 1, favors the high lattice parameter phase ($a_{H}$ = 10.665 \r{A}, $W_{H}$ = 64 mass \%) while the air-grown sample, where EPMA indicates Nd/Zr $<$ 1, favors the low lattice parameter phase ($a_{L}$ = 10.651 \r{A}, $W_{L}$ = 59 mass \%). The increased variation seen across the larger length scales of the EPMA line scans (Fig. \ref{fig:Figure4}(c,d)) suggests that, across a boule, there exists a continuous distribution of lattice parameters, rather than just two, which maps to the local Nd/Zr ratio. This likely explains why the observed set of lattice parameters is not the same between the two samples. The shift in the refined lattice parameters and weight fractions represents a change in the \textit{average} value of a continous lattice parameter distribution. This, in turn, reflects the change in the average stoichiometry. A higher level of disorder in the air-grown sample is also indicated by the higher ADP values for all sites. The A-site shows a particularly large increase ($\approx$ 43\%), consistent with the increased amount of Zr occupation.

The neutron PDF data allows us to examine the possibility of disorder within the local structure of Nd$_{2(1-x)}$Zr$_{2(1+x)}$O$_{7}$ crystals, which can be distinct from the disorder found by probing the average structure. Indeed, the poor fit to the peak in \textit{G(r)} at \textit{r} = 2.6 \r{A} when refining against the ideal pyrochlore structure suggests the need for a local structural model that goes beyond the variable position of the O1-site. The existence of a high-temperature disordered phase -- the “defect fluorite” structure (space group number 225, Fm$\bar{3}$m) -- is well known in the pyrochlore zirconates \cite{hatnean2016,subramanian1983,rushton2004,payne2013}. This provides a likely explanation for the observed local disorder since the anion sublattice of the defect fluorite structure is related to that of the pyrochlore structure by a disordering of the two distinct oxygen positions (8\textit{b} and 48\textit{f}), collapsing them to a single position (8\textit{c}). 

The inclusion of the defect fluorite phase only significantly affects $R_{W}$ in the \textit{r} = 1.5 -- 5 \r{A} range, indicating that the fluorite ``phase" is a very local disorder effect, which is best described as follows: the local environment of the Nd and Zr cations resembles a mix of pyrochlore and fluorite-type oxygen coordination (though mostly dominated by the pyrochlore-type), while the coordination of the average, crystallographic structure is that of the ideal pyrochlore. This is represented in Fig. \ref{fig:Figure2} as occupation of the oxygen 8\textit{a} interstitial which coordinates with the Zr (B) site. The change in the geometry of oxygen coordination (which forms a perfect cube for both sublattices in the fluorite case) is not illustrated. Based on the existing evidence for local disorder in polycrystalline samples \cite{blanchard2012,payne2013}, this defect fluorite character is most likely endemic to all FZ-grown samples, regardless of growth parameters.

Having demonstrated the relationship between lattice parameter, Nd/Zr ratio, and growth pressure, we now turn to the impacts of defect modes on the magnetic properties. Considering the $\chi(T)$ data for the HP-Ar grown sample, the qualitative aspects of the Curie-Weiss analysis are unaffected by the decrease in disorder, relative to air-grown samples \cite{xu2015,xu2018,hatnean2015_1}. The $\Theta_{CW}$ remains small and positive, suggesting that the balance between the FM dipolar interactions and AFM superexchange interactions is not greatly modified. Furthermore, the effective moment in this sample ($\mu_{eff}$ = 2.60 $\mu_{B}$/Nd) remains reduced relative to the expected free-ion value for Nd$^{3+}$ moments (3.62 $\mu_{B}$/Nd). However, this $\mu_{eff}$ value is $\approx$ 5\% larger than that of a previous report which used the same Curie-Weiss procedure for an air-grown sample \cite{xu2019}. As indicated in Fig. \ref{fig:Figure7}, $M_{sat}$ for the HP-Ar sample is also enhanced relative to air-grown samples \cite{hatnean2015_1,xu2019}. These results are all consistent with an increase in the amount of magnetic Nd$^{3+}$ present in crystals grown under a high-pressure Ar environment. 

The low-temperature $C_{P}$\textit{(T)} data also evince changes in the magnetism. Within our set of samples, the $C_{P}$\textit{(T)} data show $T_{AFM}$ increases in tandem with the average lattice parameter (Fig. \ref{fig:Figure9}, inset), reaching 355 mK for the HP-Ar sample. The measurement of $T_{AFM}$ by single crystal neutron scattering (Fig. \ref{fig:Figure10}(a)) -- which probes a much larger sample volume than the  $C_{P}$\textit{(T)} data -- shows that the intensity on the (220) reflection increases below 365 mK, consistent with the  $C_{P}$\textit{(T)} data given the lower density of temperatures sampled.  These values remain slightly reduced relative to powder samples ($T_{AFM}$ = 370 -- 390 mK \cite{xu2015, blote1969}), which can likely be attributed to the distribution of Nd/Zr ratios present in the sample. In comparison, previously reported, air-grown crystals also yield reduced $T_{AFM}$ values of 285 mK \cite{lhotel2015} and 310 mK \cite{opherden2017}.  These results show that the onset of long-range magnetic order is sensitive to the amount of structural disorder, which primarily manifests as a variation in the extent of negative stuffing. In contrast, the local disorder detected in the neutron PDF data does not appear to be a major factor in reducing $T_{AFM}$, as the fraction of defect fluorite coordination is comparable for the HP-Ar and the HP-80:20 samples. 

Low defect concentrations can have outsized effects on the magnetism of frustrated systems because the ground state selection is often determined by a finely tuned balance of different interactions. Yb$_{2}$Ti$_{2}$O$_{7}$ provides a particularly striking example, where varying levels of disorder, consisting of O vacancies and/or (“positive”) stuffing, was determined to be responsible for the variety of observed ground states \cite{ross2012,bowman2019}. In the present case, there is no signature of scattering intensities being redistributed away from the expected wave vectors of AIAO order, and the AIAO state in Nd$_{2}$Zr$_{2}$O$_{7}$ appears robust across an appreciable range of disorder/defect densities. 

Instead, more quantitative modifications to the ordered state, such as the increased $T_{AFM}$ in the HP-Ar sample, can be well understood by a reduction in the amount of negative stuffing. The presence of non-magnetic Zr$^{4+}$ on the A-site breaks up the Nd$^{3+}$ superexchange network, which weakens the overall  exchange interaction. This can be contrasted with disorder in Yb$_{2}$Ti$_{2}$O$_{7}$ where stuffing occurs via Yb$^{3+}$ occupying the B-site \cite{ross2012,bowman2019,baroudi2015}. In that case, stuffing provides an avenue for new magnetic exchange pathways and can qualitatively modify spin correlations, even at low defect densities \cite{ross2012,bowman2019}. That this does not happen in the case of Nd$_{2}$Zr$_{2}$O$_{7}$ is rationalized by the different type of stuffing.

We turn now to the diffuse scattering at \textit{T} = 38 mK (\textit{i.e. T} $< T_{AFM}$ ) observed for the HP-Ar crystal (Fig. \ref{fig:Figure10}(b). While the appearance of the pinch-point pattern is unaffected by the reduction in disorder, the temperature evolution appears modified relative to earlier studies on air-grown samples. Previous reports have shown that the intensity is comparable at low temperature (60 mK) and 450 mK, with a slight peak near $T_{AFM}$ and a large reduction by 750 mK \cite{petit2016}. Estimation of the magnetic moment which produces the pinch-point pattern follows a qualitatively similar trend \cite{ PhysRevLett.126.247201}. Although we do not have the data to comment on the peak at $T_{AFM}$, the pinch-point pattern in our sample persists nearly unchanged up to 800 mK (the highest temperature measured). This is shown by the integration presented in Fig. \ref{fig:Figure10}(a), where the values at 500 mK and 800 mK are almost identical. Direct comparison of the diffuse scattering heat maps (not shown) confirms this conclusion. Based on the fact that the pinch-point pattern intensity is substantially weakened by $\approx$ 750 mK in an air-grown sample \cite{petit2016} and remains robust at 800 mK in a HP-Ar sample with $\approx$ 1 \% negative stuffing, the spin-ice correlations seem sensitive to this degree of disorder. A recent report suggests that a separate spin-ice-like state may exist above $T_{AFM}$ \cite{PhysRevLett.126.247201}; however our current data are unable to address this possibility.  

The  main temperature-induced change in the diffuse scattering above $T_{AFM}$ in HP-Ar grown crystals lies in the region between the arms of the pinch-point pattern, where intensity coalesces around the (113) and (220)-type reflections as the temperature is lowered from 800 to 500 mK (Fig. \ref{fig:Figure10}(c)). Two possible origins for this diffuse scattering present themselves: 1) The onset of short-range, AIAO correlations and/or 2) scattering by monopole creation or hopping \cite{xu2020}. We cannot directly address the latter possibility since evidence for monopole dynamics comes in the form of a gapped continuum of excitations and our measurement is energy-integrated. However, assuming the former possibility, our analysis indicates an AIAO correlation length of 55 $\pm$ 10 \r{A} at 500 mK.
\FloatBarrier

\section{Conclusions \label{con}}

Using a combination of EPMA and X-ray/neutron scattering we have demonstrated that the reduced lattice parameter of air-grown Nd$_{2}$Zr$_{2}$O$_{7}$ single-crystals is due to negative stuffing (\textit{i.e}. occupation of the A-site by Zr). This occurs during FZ-growth due to preferential evaporation of Nd$_{2}$O$_{3}$, which we find can be largely mitigated by a 70 bar Ar overpressure. Neutron total scatting measurements also demonstrate the occurrence of a highly local form of disorder on the oxygen sublattice, which is only somewhat reduced via high-pressure growth. The HP-Ar grown crystals possess an optimized Nd stoichiometry with a lattice parameter approaching the polycrystalline value and a negative stuffing level of $\approx$ 1\%. However, this reduction in disorder does not qualitatively modify the underlying physics responsible for the signatures of magnetic moment fragmentation. Instead, the reduction in disorder raises the temperature range over which the AIAO order and, potentially, the diffuse pinch-point scattering are stabilized. Our results provide an important experimental demonstration of the relationship between synthesis environment, structural disorder, and magnetism in Nd$_{2}$Zr$_{2}$O$_{7}$.

\begin{acknowledgments}
We would like to thank Carlos Levi for a number of insightful discussions.  This work was partially supported by the U.S. Department of Energy, Office of Basic Energy Sciences, Division of Materials Sciences and Engineering under Award No. DE-SC0017752. A portion of this research used resources at the Spallation Neutron Source, a DOE Office of Science User Facility operated by the Oak Ridge National Laboratory. Use of the Advanced Photon Source at Argonne National Laboratory was suppoted by the U. S. Department of Energy, Office of Science, Office of Basic Energy Sciences, under Contract No. DE-AC02-06CH11357. This work made use of the MRL Shared Experimental Facilities which are supported by the MRSEC Program of the NSF under Award No. DMR 1720256, a member of the NSF-funded Materials Research Facilities Network and it also used facilities supported via the UC Santa Barbara NSF Quantum Foundry funded via the Q-AMASE-i program under award DMR-1906325. This work was also supported by the Office of Naval Research under grant N00014-19-1-2377 monitored by Dr. D.A. Shifler. Additional funding support for crystal growth was provided by the W. M. Keck Foundation. G.L. gratefully acknowledges support for this work from the National Science Foundation (NSF) through DMR 1904980 with additional support provided by Bates College internal funding.  The identification of any commercial product or trade name does not imply endorsement or recommendation by the National Institute of Standards and Technology.
\end{acknowledgments}
\bibliography{Bibliography_07122021}
\FloatBarrier
\end{document}